\documentclass[a4paper,12pt]{article}

\usepackage{graphicx}
\usepackage{epsfig}
\usepackage{amsfonts,amsmath}
\usepackage{amssymb}
\usepackage{epsf}
\usepackage{hyperref}
\usepackage{authblk}

\usepackage{color}
\usepackage[normalem]{ulem}
\usepackage[dvipsnames]{xcolor}
\usepackage{cite}

\newcommand{\hh}[1]{#1}

\title{Phase transitions of boson stars in scalar-tensor theories}

\author[1]{Hyat Huang\thanks{\href{mailto:hyat@mail.bnu.edu.cn}{hyat@mail.bnu.edu.cn}}}
\author[2]{Burkhard Kleihaus \thanks{\href{mailto:b.kleihaus@uni-oldenburg.de}{b.kleihaus@uni-oldenburg.de}}}
\author[2]{Jutta Kunz\thanks{\href{mailto:jutta.kunz@uni-oldenburg.de}{jutta.kunz@uni-oldenburg.de}}} 
\author[1]{Meng-Yun Lai\thanks{\href{mailto:mengyunlai@jxnu.edu.cn}{mengyunlai@jxnu.edu.cn}}}
\author[3]{Eugen Radu\thanks{\href{mailto:eugen.radu@ua.pt}{eugen.radu@ua.pt}}}
\author[1]{De-Cheng Zou\thanks{\href{mailto:dczou@jxnu.edu.cn}{dczou@jxnu.edu.cn}}}

\affil[1]{College of Physics and Communication Electronics, Jiangxi Normal University, Nanchang 330022, China}
\affil[2]{Institut f\"ur  Physik, Universit\"at Oldenburg, Postfach 2503,
D-26111 Oldenburg, Germany}
\affil[3]{Departamento de Matem\'atica da Universidade de Aveiro and
Centre for Research and Development in Mathematics and Applications (CIDMA),
Campus de Santiago, 3810-183 Aveiro, Portugal}

\date{\today}

\begin{document}

\maketitle

\begin{abstract}
In scalar-tensor theories, compact objects may experience spontaneous scalarization.
Recently, it was shown that matter-induced spontaneous scalarization of neutron stars is predominantly associated with a first-order phase transition.
Here we consider matter-induced spontaneous scalarization of boson stars.
Employing a repulsive quartic potential for the bosonic matter, we find only first-order phase transitions.
\end{abstract}

\section{Introduction}

While General Relativity (GR) is well tested in the weak gravity regime \cite{Will:2005va,Will:2018bme}, the strong gravity regime is far less constrained, leaving considerable freedom for alternative theories of gravity \cite{Faraoni:2010pgm,Berti:2015itd,CANTATA:2021ktz}.
Among the plethora of alternative theories suggested, the scalar-tensor theories (STTs) represent very early and natural generalizations \cite{Jordan:1949zz,Fierz:1956zz,Jordan:1959eg,Brans:1961sx,Dicke:1961gz}.
These STTs feature besides the metric an additional scalar field that is mediating the gravitational interaction together with the metric.

Compact objects like black holes and neutron stars are excellent laboratories to test and constrain alternative theories of gravity in the strong gravity regime, since their properties may differ considerably in alternative theories and in GR.
STTs, for instance, allow for spontaneous scalarization of neutron stars, when an appropriate coupling function for the gravitational scalar field is chosen.
In that case, the neutron star solutions of GR remain solutions of the STT, but a tachyonic instability gives rise to scalarized solutions, when the matter is sufficiently compact \cite{Damour:1993hw,Doneva:2013qva}.
For rotating neutron stars, the deviations
from the GR solutions are even
significantly enhanced compared to the static case.

Spontaneous scalarization of the vacuum black holes of GR can, in contrast, be triggered by sufficiently strong curvature in the presence of higher-order curvature invariants in the action of alternative gravity theories (see e.g. \cite{Doneva:2022ewd}).
For rotating black holes, even two distinct mechanisms for spontaneous scalarization arise that are associated with different signs of the coupling constant \cite{Dima:2020yac,Hod:2020jjy,Herdeiro:2020wei,Berti:2020kgk}.
When employing the same sign as in the static case, rotation leads to a suppression of scalarization and thus a decrease of the domain of existence of scalarized black holes.
When employing the opposite sign, however, scalarization sets in at sufficiently high angular momentum.

Besides neutron stars and black holes, also hypothetical compact objects like boson stars \cite{Jetzer:1991jr,Lee:1991ax,Schunck:2003kk,Liebling:2012fv} may experience spontaneous scalarization \cite{Torres:1997np,Whinnett:1999sc,Alcubierre:2010ea,Ruiz:2012jt,Kleihaus:2015iea,Evstafyeva:2023kfg}.
Indeed, the complex scalar field constituting the star undergoes matter-induced spontaneous scalarization in STTs similar to neutron stars.
Analogously to neutron stars, also the presence of rotation significantly enhances the deviations from the GR case \cite{Kleihaus:2015iea}.
Moreover, this spontaneous scalarization carries over to Kerr black holes with scalar hair \cite{Herdeiro:2014goa}, which then feature also a gravitational scalar field in addition to the complex scalar hair \cite{Kleihaus:2015iea}.

Recently, the nature of the phase transition between neutron stars in GR and their scalarized brethren was analyzed \cite{Unluturk:2025zie,Muniz:2025egq}.
This led to the somewhat surprising result that a first-order phase transition is far more common than a second-order one, when the severe observational constraints on the STT are evaded by allowing for a small mass of the gravitational scalar field \cite{Freire:2024adf}.
Here we therefore revisit boson stars in STT in order to study the nature of their phase transition.
For the boson stars, we employ either mostly a repulsive 4th-order potential \cite{Colpi:1986ye}, but allow also for mini-boson stars, where the self-interaction coupling constant vanishes.
Surprisingly, we then see only first-order transitions in our calculations.
 
The paper is organized as follows: 
In the next section, we recall the theoretical setting for STTs giving rise to spontaneous scalarization and to boson stars.
We present our results in section 3, first for spherically symmetric boson stars and then for rotating boson stars.
We conclude in section 4.

\section{Theoretical setting}

We here follow largely our previous paper on boson stars in STTs \cite{Kleihaus:2015iea}, considering slightly more general theories by allowing for a mass of the gravitational scalar field, while applying the same notation and methods.

\subsection{Scalar-tensor theories}

We start from the physical Jordan frame, where the action is given by
\begin{eqnarray} \label{JFA}
S &=& \frac{1}{16\pi G_{*}} \int d^4x \sqrt{-{\tilde
g}}\left({F(\Phi)\tilde R} - Z(\Phi){\tilde
g}^{\mu\nu}\partial_{\mu}\Phi
\partial_{\nu}\Phi   -2 W(\Phi) \right) \nonumber \\[2ex]
&+&
S_{m}\left[\Psi_{m};{\tilde g}_{\mu\nu}\right] ,
\end{eqnarray}
where the first term represents the gravitational action with the bare gravitational constant $G_{*}$, the spacetime metric ${\tilde g}_{\mu\nu}$, the Ricci scalar curvature ${\tilde R}$, and the real gravitational scalar field $\Phi$, whereas the second term 
denotes the action of the matter fields, i.e., the complex scalar field $\Psi_m$.

The gravitational action contains the functions $F(\Phi)$, $Z(\Phi)$ and $W(\Phi)$, that are subject to several physical restrictions.
The function $F$ is required to be positive, $F(\Phi)>0$, so that gravitons have positive energy.
Furthermore, the kinetic energy of the scalar field should be non-negative, which requires the constraint
\begin{equation}
2F(\Phi)Z(\Phi) + 3[dF(\Phi)/d\Phi]^2 \ge 0.
\end{equation}

In the Jordan frame the matter action $S_{m}$ depends only on the space-time metric ${\tilde g}_{\mu\nu}$ and the minimally coupled complex scalar field $\Psi_{m}$.
An additional coupling to the gravitational scalar field $\Phi$ would violate the weak equivalence principle.

We now turn to the Einstein frame which we later employ to construct the boson star solutions.
The Einstein frame is conformally related to the Jordan frame and has metric $g_{\mu\nu}$
\begin{equation}\label {CONF1}
g_{\mu\nu} = F(\Phi){\tilde g}_{\mu\nu} .
\end{equation}
The action in the Einstein frame can be expressed simply by introducing a new scalar field $\phi$, defined via 
\begin{equation}\label {CONF2}
\left(\frac{d\phi}{d\Phi} \right)^2 = \frac{3}{4}\left(\frac{d\ln(F(\Phi))}{d\Phi } \right)^2 + \frac{Z(\Phi)}{2 F(\Phi)} ,
\end{equation}
and the new functions
\begin{equation}\label{CONF3}
{\cal A}(\phi) = F^{-1/2}(\Phi) \,\,\, ,\nonumber \\
2V(\phi) = W(\Phi)F^{-2}(\Phi) .
\end{equation}
The action in the Einstein frame then reads
\begin{eqnarray}
S= \frac{1}{16\pi G_{*}}\int d^4x \sqrt{-g} \left(R -
2g^{\mu\nu}\partial_{\mu}\phi \partial_{\nu}\phi -
4V(\phi)\right)+ S_{m}[\Psi_{m}; {\cal A}^{2}(\phi)g_{\mu\nu}] ,
\end{eqnarray}
where the Ricci scalar $R$ is defined in terms of the metric in the Einstein frame $g_{\mu\nu}$.

Variation of the action in the Einstein frame with respect to the metric in the Einstein frame $g_{\mu\nu}$, the new scalar field $\phi$, and the complex scalar field $\Psi_m$
leads to the coupled set of field equations in the Einstein frame.

Since we are interested in matter-induced spontaneous scalarization of boson stars, we employ an appropriate coupling function ${\cal A}(\phi)$ \cite{Damour:1993hw,Kleihaus:2015iea}
\begin{equation}
\ln {\cal A}(\phi) = \frac{1}{2} \beta \phi^2  ,
\label{funA}
\end{equation}
where $\beta$ is a parameter.
Furthermore, we either choose for the potential of the STT $V(\phi)=0$, corresponding to a massless scalar field $\phi$, or we choose a small mass $m_\phi$ for the scalar field $\phi$ by selecting the potential
\begin{equation}
    V(\phi)= \frac{1}{2} m_\phi^2 \phi^2 .
\end{equation}

In the Einstein frame the action $S_{m}[\Psi_{m}; {\cal A}^{2}(\phi)g_{\mu\nu}]$ for the complex scalar field $\Psi$ reads
\begin{equation}\label{Smat}
S_{m}\left[\Psi_{m};{\tilde g}_{\mu\nu}\right]
= - \int d^4x \sqrt{-g} 
\left[ \frac{1}{2} {\cal A}^2(\phi)  g^{\mu\nu}
\left( \Psi_{, \, \mu}^* \Psi_{, \, \nu} 
+ \Psi _ {, \, \nu}^* \Psi _{, \, \mu} \right) 
+ {\cal A}^4(\phi)  U( \left| \Psi \right|) \right] ,
\end{equation}
where we have included a self-interaction potential $U( \left| \Psi \right|)$ for the complex scalar field $\Psi$,
\begin{equation}\label{SmatU}
U( \left| \Psi \right|) =
m_b^2 \left| \Psi \right|^2 + \Lambda \left| \Psi \right|^4 \ ,
\end{equation}
with the boson mass $m_b$ and the coupling constant $\Lambda>0$.

\subsection{Boson stars}

To construct rotating boson star solutions, we employ the stationary axially symmetric line element 
\cite{Kleihaus:2000kg,Kleihaus:2004gm,Kleihaus:2015iea}.
\begin{equation}
ds^2 = - f_0 dt^2 + \frac{f_1}{f_0} \left( f_2 \left[
dr^2 + r^2 d \theta^2 \right]
+ r^2 \sin^2 \theta \left[ d\varphi - f_3 dt \right]^2 \right) ,
\label{metric}
\end{equation}
with the functions $f_i(r,\theta)$, $i=0,...,3$.
In the static case, the boson stars are spherically symmetric and the line element simplifies accordingly, i.e., $f_3=0$ and $f_2=1$.
The gravitational scalar field $\phi$ is parametrized in both cases as $\phi(r,\theta)$ and $\phi(r)$, respectively.
As usual, the Ansatz for the complex scalar field $\Psi$ is taken to be
\begin{equation}
\Psi = \psi(r,\theta) \, e^{i \omega t + i n \varphi}
\end{equation}
with the real function $\psi$, and the boson frequency $\omega$.
The integer $n$ provides single-valuedness for the complex scalar field,
with $n=0$ in the static case and $n>0$ in the rotating case.
Because of the time dependence and azimuthal angle dependence of the complex scalar field, the latter has fewer symmetries than the space-time metric.

In the Lagrangian for the boson field $\Psi$ 
we employ a quartic self-interaction potential (\ref{SmatU})
with coupling constant $\Lambda$.

%
\hh{
The boundary conditions at the origin are obtained by requiring regularity,
i.~e.~all functions possess a Taylor expansion in terms of cartesian coordinates.
This is guaranteed if 
\begin{equation}
\partial_r f_i(0,\theta) = 0 , \ \ \ i=0,1,2,3  \ ,  \ \ \
 \psi(0,\theta) = 0 \ , \ \ \ \partial_r \phi(0,\theta) = 0 . 
\end{equation}
Similarly, elementary flatness and regularity on the symmetry axis requires
\begin{equation}
 \partial_\theta f_i(r,0) = 0 , \ \ \ i=0,1,3 \ , \ \ \ f_2(r,0) =1 \ ,  \ \ \
 \psi(r,0) = 0 \ , \ \ \partial_\theta \phi(r,0) = 0 .
\end{equation}
At spatial infinity we require vacuum and Minkowski spacetime, i.~e.
\begin{equation}
f_i(\infty,\theta) = 1, \ \ \ i=0,1,2 \ , \ \ \ f_3(\infty,\theta) =0 \ ,  \ \ \
 \psi(\infty,\theta) =\phi(\infty,\theta) = 0 .
\end{equation}
Finally, we here consider only solutions with reflection symmetry with respect to the equatorial plane,
i.~e.~even functions in $\cos\theta$ at $\theta=\pi/2$,
\begin{equation}
\partial_\theta f_i(r,\pi/2) = 0 , \ \ \ i=0,1,2,3 \ ,  \ \ \
 \partial_\theta \psi(r,\pi/2) =  \partial_\theta \phi(r,\pi/2) = 0 \ .
\end{equation}
Note that in the spherically symmetric case $n=0$ the boundary conditions for $\psi$ are different at the center and on the symmetry axis, 
i.~e.~$\partial_r \psi(0,\theta) =0$ and $\partial_\theta \psi(r,0)=0$.
This is due to the absence of the factor $e^{i n \varphi}$ in the complex scalar field $\Psi$.
}

Turning now to the global charges of the boson star solutions, we can read off their mass $M$ and angular momentum $J$ from the asymptotic expansion of the
metric functions $f_0$ and $f_3$ as follows,
\begin{eqnarray} 
f_0 &=& 1 -2 M/r + O(r^{-2})\ , \\
f_3 &=&  2 J/r^3 + O(r^{-4})\ . 
\end{eqnarray} 
Since the space-time is asymptotically flat, the expansion yields the charges not only for GR but also for the STTs.

The third global charge is the Noether charge $N$ of the boson stars, that arises due to the global phase invariance of the action.
The transformation of the complex scalar field
\begin{equation}
    \Psi \rightarrow \Psi e^{i\chi}
\end{equation}
(with $\chi$ a real constant)
leads to the conserved current
\begin{eqnarray}
j^{\mu} & = &  - i {\cal A}^2(\phi)\left( \Psi^* \partial^{\mu} \Psi 
 - \Psi \partial^{\mu}\Psi ^* \right) \ , \ \ \
j^{\mu} _{\ ; \, \mu}  =  0 \ .
\end{eqnarray}
The Noether charge $N$ 
\begin{equation}
N = \int_{\Sigma} \sqrt{ - g} j^0 
d\bar{r} d\theta d\varphi 
\label{Q}
\end{equation}
then represents the particle number of the boson stars.
For rotating boson stars the relation $J=nN$ holds.

\subsection{Description of phase transitions}

The spontaneous scalarization of compact objects can be interpreted as a phase transition between two equilibrium configurations of the same system: 
an unscalarized one, corresponding to the GR solution with vanishing gravitational scalar field, and a scalarized one with a non-vanishing scalar field. 
Depending on the underlying parameters of the STT, this transition may proceed either continuously or discontinuously. 

As pointed out by Ünlütürk \textit{et al.}~\cite{Unluturk:2025zie}, the onset of scalarization can be phenomenologically modeled within a Landau-type framework, in which the total energy (or ADM mass) of a compact object is expressed as an even polynomial in a suitable order parameter~$Q$ (or the central value $\psi_c$) representing the strength of the scalar field. 
For a given baryon mass~$M_b$, one may expand the ADM mass as
\begin{equation}
M_{\mathrm{ADM}} = M_0(M_b)
     + a(M_b) Q^2
     + \frac{1}{2} b(M_b) Q^4
     + \frac{1}{3} c(M_b) Q^6 + \dots.
\label{eq:landau}
\end{equation}
Note that only even terms of $Q$ arise here because of the symmetry of $\phi \to -\phi$. 

In this picture, the stable equilibrium configurations correspond to local minima of $M_{\mathrm{ADM}}(Q)$. 
When $b > 0$ and $a(M_b)$ changes sign from positive to negative at some critical baryon mass $M_{\mathrm{crit}}$, the minimum of the potential shifts continuously from $Q=0$ to $Q \neq 0$. 
This represents a second-order (continuous) phase transition, where the scalar charge and other physical quantities vary smoothly across the critical point. 
When $b < 0$ and $c > 0$, however, the potential develops two distinct local minima separated by a barrier, so that both scalarized ($Q \neq 0$) and unscalarized ($Q=0$) configurations can coexist for a finite interval of $M_b$. 
In this regime, the system exhibits metastability, and the transition between the two phases involves a discontinuous jump in the order parameter $Q$. 
The point where the two minima become degenerate defines the actual transition point, marking a first-order phase transition.

These general arguments provide a qualitative understanding of how the order of scalarization depends on the microscopic parameters of the theory, and how metastable branches naturally emerge in the presence of multiple local minima of the energy functional.

For boson stars, we employ an analogous energetic criterion to determine the order of the transition between the GR branch and the scalarized branch. 
Since the total boson mass $M_b$ is conserved, the relevant thermodynamic potential is the total binding energy,
\begin{equation}
E_{\mathrm{bind}} = \frac{M_b - M_{\mathrm{ADM}}}{M_b}.
\label{eq:binding}
\end{equation}
A second-order transition would manifest as a smooth bifurcation of the scalarized branch from the GR branch, with the scalar field amplitude growing continuously from zero and no overlap of energetically stable configurations. 
In contrast, a first-order transition is identified by the following features.
Firstly, the GR and scalarized branches coexist over a finite range of $M_b$.
{Here two} 
branches correspond to locally stable equilibria, separated by an intermediate unstable configuration. 
The binding energy curves $E_{\mathrm{bind}}(M_b)$ 
{of} the two stable branches 
{then} intersect, 
{and} the energetically favored configuration switches discontinuously from the GR solution (for lower $M_b$) to the scalarized solution (for higher $M_b$) at the crossing point. This pattern is entirely analogous to that found for neutron stars in Ref.~\cite{Unluturk:2025zie}.

\section{Results}

In our previous calculations \cite{Kleihaus:2015iea}, we employed the STT parameter $\beta=-4.7$ with a vanishing mass of the gravitational scalar, $m_\phi=0$, satisfying observational constraints for the massless case.
At the same time, we chose for the self-interaction coupling constant mainly the value $\Lambda=300$, but also performed calculations for other values of $\Lambda$.
Here we supplement these calculations by using the parameter $\beta=-10$ together with the masses $m_\phi=0$ and $m_\phi/m_b=10^{-3}$, but we also allow for further values of $\beta$.
Moreover, we consider a large range of values for the self-interaction parameter $\Lambda$, including the case $\Lambda=0$, i.e., mini-boson stars.

To present our results in dimensionless units, we introduce the quantities $M_0$, $N_0$, and $\omega_0$
\begin{equation}
M_0 = m_{Pl}^2/m_b \ , ~N_0= m_{Pl}^2/m_b^2 \ , ~
\omega_0=m_b \ ,
\end{equation}
where $m_{Pl}$ is the Planck mass.
If we choose $M_0 = M_{\odot}$ we get $m_b = 1.335\times 10^{-10} $ eV and $m_\phi = 1.335\times 10^{-13} $ eV. 
This fits in the allowed range \cite{Yazadjiev:2016pcb}:
$10^{-16} {\rm eV} <  m_\phi < 10^{-9} {\rm eV}$.

As previously, the system of non-linear coupled partial differential equations has been solved subject to the given set of boundary conditions.
The equations have been discretized on a non-equidistant grid in the compactified radial coordinate $x=\frac{r}{r+1}$ and the polar angle $\theta$. 
We have used typical grids of sizes 
$251 \times 30$
with integration regions $0 \le x \le 1$ and $0 \le \theta \le \pi/2$. 
To obtain the solutions, we have used the professional package FIDISOL/CADSOL \cite{FIDISOL}, that employs a Newton-Raphson method and provides an error calculation.

\hh{
In more detail, the exact solution is split into an approximate solution and a small correction. 
The discretized PDEs are expanded up to first order in the correction. 
This yields a linear equation for the correction, where the inhomogeneous part consists of the discretized PDEs evaluated with the approximate solution. 
The linear equation is solved for the correction which is subsequently added to the approximate solution.
The result is an improved approximate solution, which serves as starting point for the next iteration step. 
In this way corrections are computed successively, until the inhomogeneous part becomes arbitrarily small.
}

\hh{
Although the discretized PDEs are solved (almost) exactly, there is still a numerical error.
This is due to the fact that the discretized PDEs are only an approximation of the exact PDEs.
To obtain an estimate of the numerical error, solutions obtained with different orders of consistency of the discretization scheme are compared.
Typically the maximal relative error is below $10^{-5}$ for the rotating boson stars and even less for the non-rotating ones.
}

\hh{
The boundary conditions are treated formally as part of the PDEs and solved during the Newton-Raphson iterations as well.
}

In the following, we present the results first for the non-rotating case and then for the rotating case.
We note already that the results with a small mass of the gravitational scalar are quite similar to the massless case, when the other parameters are kept fixed.\\

\subsection{Spherically symmetric boson stars \boldmath $(n=0)$ \unboldmath }

\begin{figure}[t!]
\begin{center}
\vspace{-0.5cm}
\mbox{
(a)\hspace*{-0.5cm}\includegraphics[height=.225\textheight, angle =0]{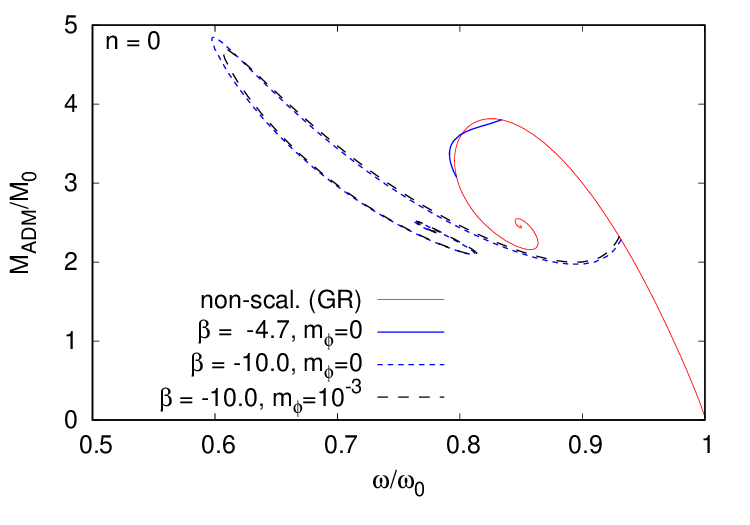}
(b)\hspace*{-0.5cm}\includegraphics[height=.225\textheight, angle =0]{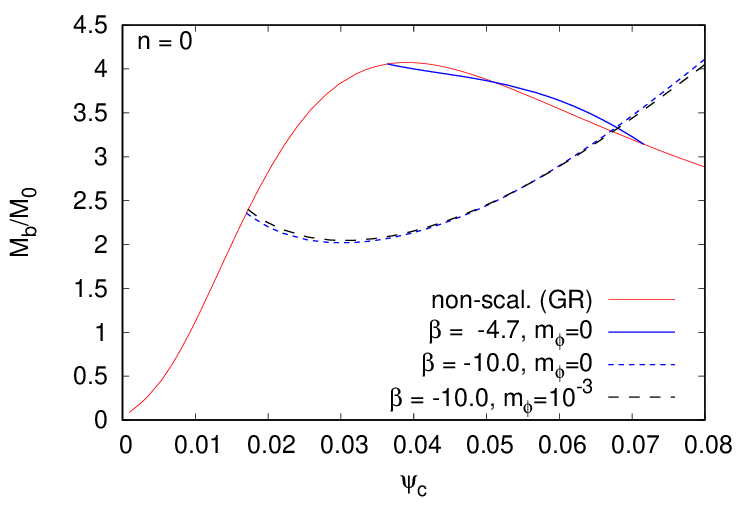}
}
\end{center}
\vspace{-0.5cm}
\caption{Non-rotating boson stars ($n=0$): 
(a) The scaled ADM mass $M_\text{ADM}/M_0$ versus the scaled boson frequency $\omega/\omega_0$ is shown for non-scalarized (GR) boson stars (solid) with self-interaction coupling constant $\Lambda=300$,
and for scalarized boson stars for a massless ($\beta=-4.7$ and $\beta=-10$, dotted), resp.~a massive ($m_\phi/m_b=10^{-3}$, $\beta=-10$, dashed) gravitational scalar field; 
(b) analogous to (a) for the scaled boson mass $M_b/M_0$ versus the central value $\psi_c$ of the boson field.
\label{fig1}
}
\end{figure}

We start by briefly recalling the non-rotating spherically symmetric boson star solutions in GR, focusing on mini-boson stars and boson stars with a repulsive potential for the complex scalar field.
The boson star solutions emerge from the vacuum at the maximal frequency $\omega=m_b$ and then monotonically grow in mass with decreasing frequency until a maximum of the mass is reached, as seen in Fig.~\ref{fig1}(a).
Beyond this maximum, the solutions start a spiraling behavior in this type of diagram and are no longer stable, analogous to neutron stars. 

Next we address the scalarized boson star solutions of this model.
Figure \ref{fig1}(a) includes scalarized boson stars for a massless gravitational scalar field for the couplings $\beta=-4.7$ and $\beta=-10$, and in addition for a massive gravitational scalar field with the scaled mass $m_\phi/m_b=10^{-3}$ for the coupling $\beta=-10$.
For $\beta=-4.7$, the scalarization starts only in the vicinity of the maximum mass and is then present in the unstable region.
So from a physical point of view, these scalarized solutions are not expected to be very relevant.

When choosing a smaller value for $\beta$, e.g., $\beta=-10$ as in the figure, scalarization sets in earlier, i.e., already on the stable GR branch.
The scalarized branch then reaches considerably higher masses and lower frequencies than the GR branch, and thus differs significantly from the GR branch. 
Note that this scalarized branch does not reconnect to the GR solutions in a second bifurcation 
but forms a spiral-like structure.
Of course, because of observational constraints, the gravitational scalar field should carry some mass.
But a small mass like the value $m_\phi/m_b=10^{-3}$ employed in the figure alters the set of scalarized boson stars only slightly as compared to the massless set.

This is also seen in Fig.~\ref{fig1}(b), where we exhibit the scaled boson mass $M_b/M_0$ versus the central value $\psi_c$ of the boson field  for the same sets of solutions.
Here the boson mass $M_b$ simply corresponds to the mass of $N$ free bosons,
\begin{equation}
 M_b = m_b N .
\end{equation}
The figure shows that for the same number of bosons there are several configurations of potential physical interest.
This is analogous to the case of neutron stars, where the physical relevance of the solutions together with the order of the phase transition was analyzed before \cite{Unluturk:2025zie,Muniz:2025egq}.

\begin{figure}[t!]
\begin{center}
\vspace{-0.5cm}
\mbox{
(a)\hspace*{-0.5cm}\includegraphics[height=.225\textheight, angle =0]{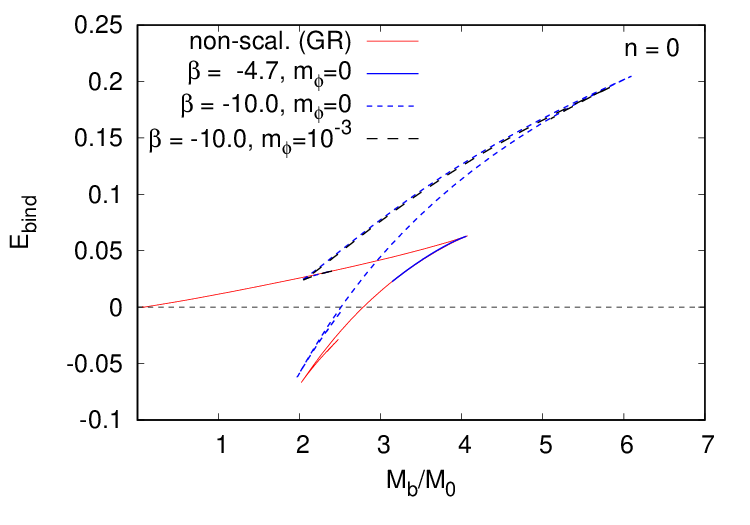}
(b)\hspace*{-0.5cm}\includegraphics[height=.225\textheight, angle =0]{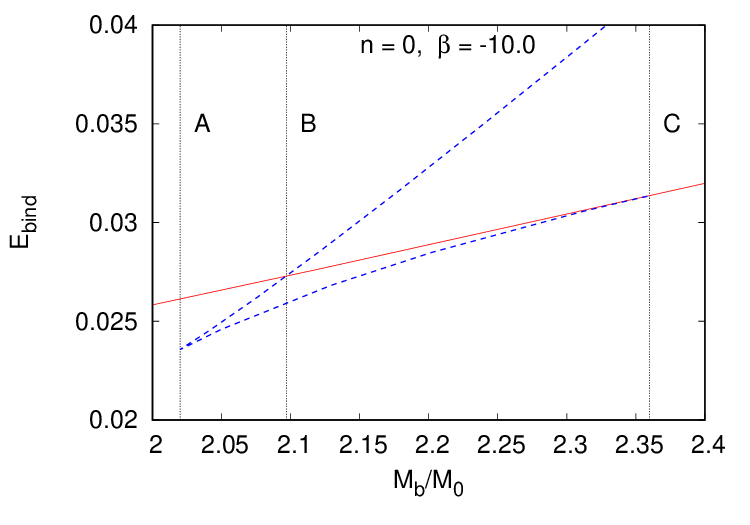}
}
\end{center}
\vspace{-0.5cm}
\caption{Non-rotating boson stars ($n=0$): 
(a) The relative binding energy $E_\text{bind}$ versus the scaled boson mass $M_b/M_0$ for the same sets of solutions as in Fig.~\ref{fig1}.
(b) Enlargement of the region close to the bifurcation 
for $\beta=-10$, $m_\phi/m_b=0$.
\label{fig2}
}
\end{figure}

Following \"Unl\"ut\"urk et al.~\cite{Unluturk:2025zie}
let us therefore consider the relative binding energy $E_\text{bind}$ 
\hh{Eq.~(\ref{eq:binding})}
as depicted in Fig.~\ref{fig2}(a) for the same sets of solutions.
The relative binding energy of the scalarized set of solutions for $\beta=-4.7$ is almost identical to that of the GR solutions for a given boson mass $M_b$.
But for $\beta=-10$, the binding energy $E_\text{bind}$ first decreases slightly below the GR values, and then crosses and rapidly exceeds the GR values. 

In Fig.~\ref{fig2}(b) we zoom into the critical region close to the bifurcation of GR boson stars and scalarized boson stars ($\beta=-10$, $m_\phi/m_b=0$).
The bifurcation point is labeled C in the figures, and the cusp of the scalarized solutions featuring the lowest boson mass is labeled A.
Between A and C, two scalarized solutions exist besides the GR solution.
The scalarized solutions with the lowest binding energy, forming the branch AC, are unstable, while the scalarized solutions with higher binding energy cross the GR solutions at B.
These scalarized solutions have lower binding energy than the GR solutions in the interval AB, but higher binding energy in the interval BC and beyond.
This makes the GR solutions energetically preferred up to B, while subsequently the scalarized solutions are preferred.

Clearly, the pattern coincides precisely with the one discussed for neutron stars \cite{Unluturk:2025zie}, making the phase transition from GR to STT boson stars a first-order phase transition.
However, so far we have shown that the phase transition is of first order only for $\beta=-10$, $m_\phi/m_b=0$, and $\Lambda=300$.
Clearly, a small mass of the gravitational scalar on the order of $m_\phi/m_b=10^{-3}$ does not change anything significantly in the above analysis.
But we should address the dependence of the order of the phase transition on the STT parameter $\beta$ and the self-coupling constant $\Lambda$ of the complex scalar field. 

\begin{figure}[t!]
\begin{center}
\vspace{-0.5cm}
\mbox{
(a)\hspace*{-0.5cm}\includegraphics[height=.225\textheight, angle =0]{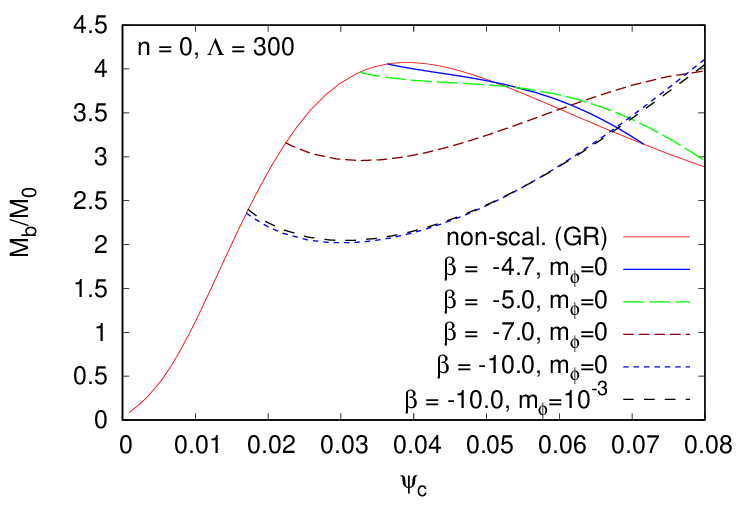}
(b)\hspace*{-0.5cm}\includegraphics[height=.225\textheight, angle =0]{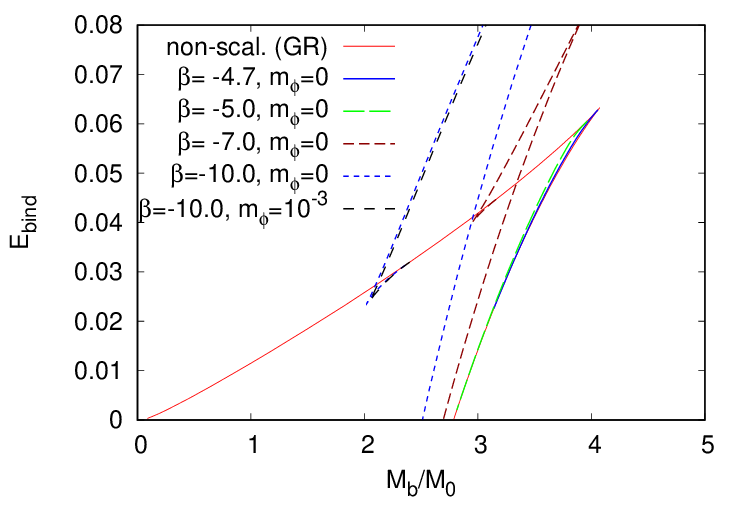}
}
\mbox{
(c)\hspace*{-0.5cm}\includegraphics[height=.225\textheight, angle =0]{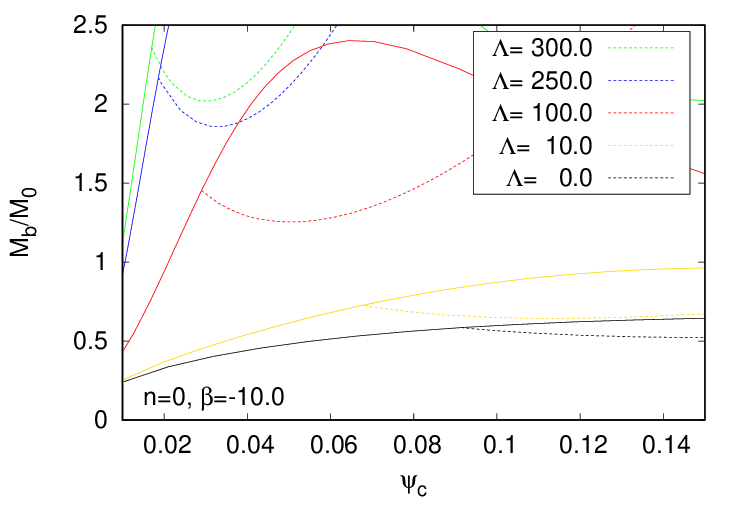}
(d)\hspace*{-0.5cm}\includegraphics[height=.225\textheight, angle =0]{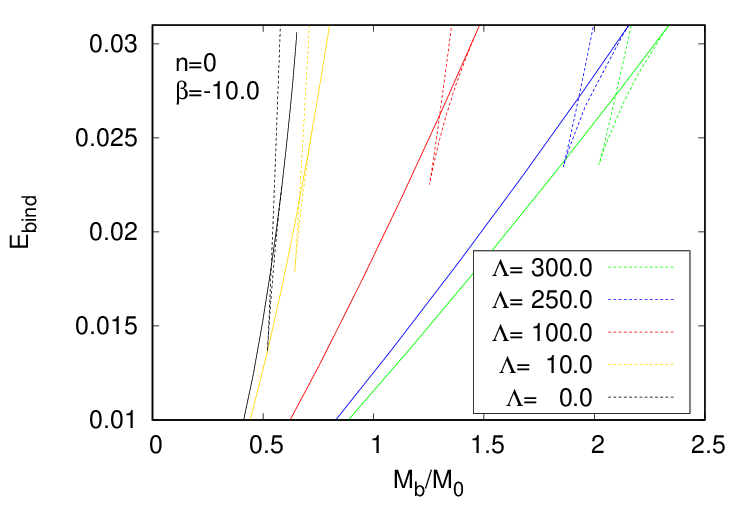}
}
\end{center}
\vspace{-0.5cm}
\caption{Non-rotating boson stars ($n=0$):
(a) The scaled boson mass $M_b/M_0$  versus the central value $\psi_c$ of the boson field is shown for the non-scalarized (GR) boson stars, and for the scalarized boson stars for a massless gravitational scalar field and a set of values of the SST parameter $\beta$ for fixed self-coupling constant $\Lambda=300$.
\hh{
(b) The relative binding energy $E_\text{bind}$ versus the scaled boson mass $M_b/M_0$ for the same sets of solutions as in (a).}
(c) Analogous to (a) for a set of values of $\Lambda$ and fixed $\beta=-10$. 
\hh{
(d) Analogous to (b) for the solutions in (c) restricted to the bifurcation regions.}
\label{fig3}
}
\end{figure}

Therefore, we now vary these two parameters to see whether they affect the order of the phase transition.
We exhibit in Fig.~\ref{fig3} the scaled boson mass $M_b/M_0$ versus the central value $\psi_c$ of the boson field 
\hh{
as well as the binding energy $E_\text{bind}$ versus the scaled boson mass $M_b/M_0$ for} 
and varying parameters.
Figures \ref{fig3}(a) \hh{and (b)} keep the self-coupling constant $\Lambda$ fixed while varying the STT parameter $\beta$, whereas Figs.~\ref{fig3}(c) \hh{and (d)} keep $\beta$ fixed while $\Lambda$ is varied.
In both cases, the scalarized solutions possess smaller boson masses after the bifurcation from the GR branch, reach a minimum mass, and then exhibit again increasing masses,
\hh{
while their binding energy features a triple branch structure, unless $\beta$ is chosen too small.}
Thus, they show the characteristic features of first-order transitions, as detailed above 
\hh{
(or no phase transition for small $\beta$)}.
The fact that we have not seen any other behavior lets us conjecture that this type of boson stars including the mini-boson stars feature only first-order transitions.

\subsection{Rotating boson stars \boldmath $(n=1)$ \unboldmath }

We now turn to rotating boson stars with rotational parameter $n=1$, i.e., these represent the families with the lowest finite angular momentum for a given particle number, since $J=nN$.
As in the spherically symmetric case, we here consider the sets of boson stars for the STT parameters $\beta=-4.7$ and $\beta = - 10$ and boson self-coupling constant $\Lambda=300$, but we leave the gravitational scalar massless.

\begin{figure}[t!]
\begin{center}
\vspace{-0.5cm}
\mbox{
(a) \hspace*{-0.5cm}\includegraphics[height=.225\textheight, angle =0]{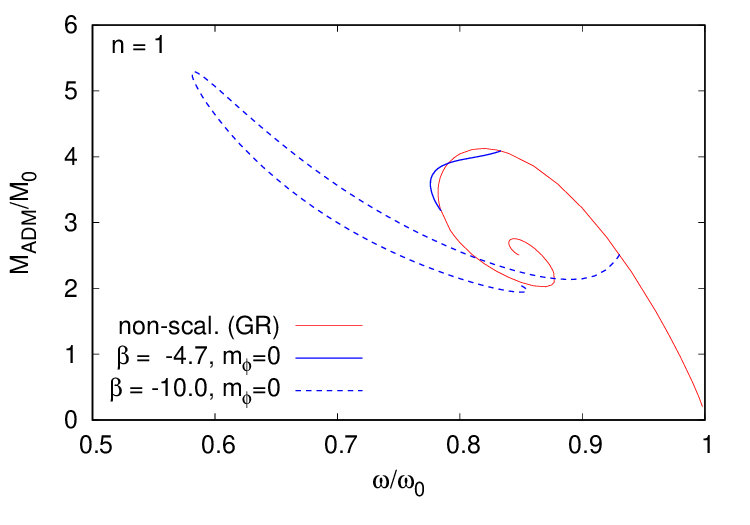}
(b)\hspace*{-0.5cm}\includegraphics[height=.225\textheight, angle =0]{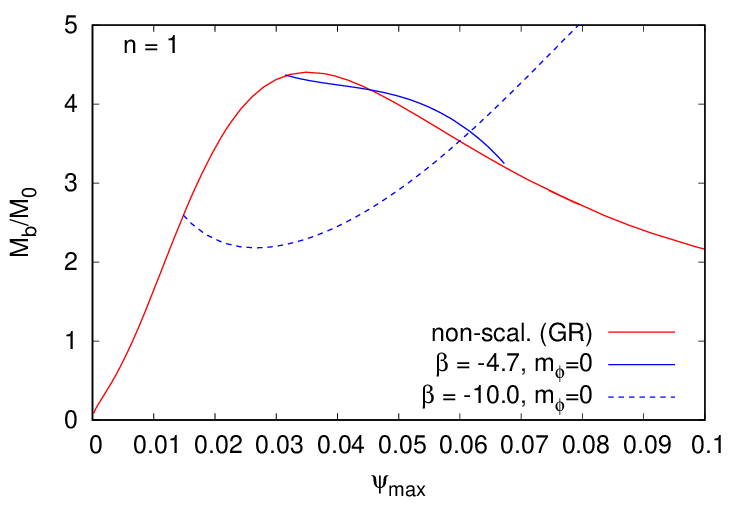}
}
\end{center}
\vspace{-0.5cm}
\caption{Rotating boson stars ($n=1$): 
(a) The scaled ADM mass $M_\text{ADM}/M_0$ versus the scaled boson frequency $\omega/\omega_0$ is shown for non-scalarized (GR) boson stars (solid) with self-interaction coupling constant $\Lambda=300$,
and for scalarized boson stars for a massless ($\beta=-4.7$ and $\beta=-10$, dotted) gravitational scalar field; 
(b) analogous to (a) for the scaled boson mass $M_b/M_0$ versus the maximal value $\psi_\text{max}$ of the boson field.
\label{fig4}
}
\end{figure}

A basic difference between the spherically symmetric non-rotating boson stars and the axially symmetric rotating boson stars is the distribution of their energy density and thus their boson field distribution.
Typically, the configurations become oblate in the presence of rotation.
In the case of boson stars, however, the energy density becomes torus-like. 
Thus the boson field assumes its maximum value $\psi_\text{max}$ on a ring in the equatorial plane
(at least for rotating boson stars with reflection symmetry \cite{Schunck:1996,Schunck:1996he,Ryan:1996nk,Yoshida:1997qf,Schunck:1999pm,Kleihaus:2005me,Kleihaus:2007vk}).

However, when we repeat the above considerations for the case of rotating boson stars, we find basically the same pattern as in the nonrotating case.
We exhibit the scaled ADM mass versus the scaled boson frequency in Fig.~\ref{fig4}(a).
As for the non-rotating solutions, the GR solutions and the scalarized solutions differ only slightly for $\beta=-4.7$, whereas for $\beta=-10$ the scalarized solutions reach considerably higher masses and lower boson frequencies.
Again this scalarized branch does not reconnect at a second bifurcation to the GR solutions.
We show in Fig.~\ref{fig4}(b) the scaled boson mass versus the maximal value $\psi_\text{max}$ of the boson field.
Again, the picture is analogous to the non-rotating case, indicating a first-order phase transition for the physically relevant solutions for $\beta=-10$.

\begin{figure}[h!]
\begin{center}
\vspace{-0.5cm}
\mbox{
(a) \hspace*{-0.5cm}\includegraphics[height=.225\textheight, angle =0]{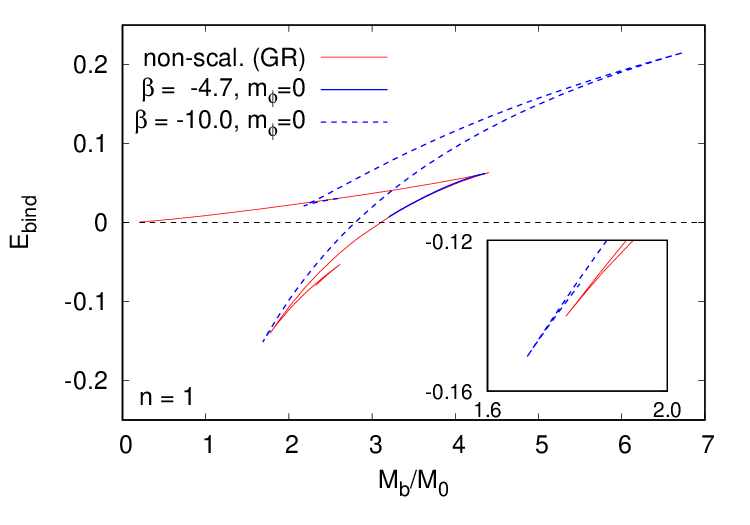}
(b)\hspace*{-0.5cm}\includegraphics[height=.225\textheight, angle =0]{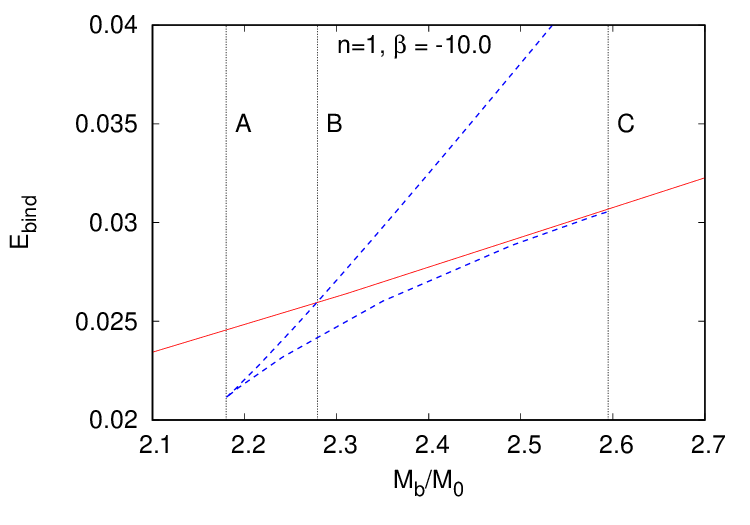}
}
\end{center}
\vspace{-0.5cm}
\caption{Rotating boson stars ($n=1$): 
(a) The relative binding energy $E_\text{bind}$ versus the scaled boson mass $M_b/M_0$ for the same sets of solutions as in Fig.~\ref{fig4}.
(b) Enlargement of the region close to the bifurcation for $\beta=-10$, $m_\phi/m_b=0$.
\label{fig5}
}
\end{figure}

This is demonstrated in detail in Fig.~\ref{fig5}, where we show the relative binding energy $E_\text{bind}$ versus the scaled boson mass.
Figure \ref{fig5}(a) gives the larger picture, whereas Fig.~\ref{fig5}(b) zooms again into the bifurcation region, indicating the cusp A, the point of equal binding energy B, and the bifurcation point C.
As before, below B GR solutions are energetically favored, whereas above B scalarized solutions on the upper branch are favored.

\section{Conclusions and Further Remarks}

Following recent investigations on the nature of the spontaneous scalarization phase transition of neutron stars in General Relativity and scalar-tensor theories \cite{Unluturk:2025zie,Muniz:2025egq}, we here have considered this question for boson stars.
We have employed a repulsive potential for the boson field, but also allowed the self-interaction of the boson to vanish, leading to mini-boson stars.
We typically observe a bifurcation of GR and STT boson stars on the physically relevant GR branch below the maximum mass, where the scalarized boson stars possess decreasing ADM mass and boson mass while the GR solutions have increasing mass.
Since the STT solutions then reach a minimum mass before the mass increases again, a cusp appears in the binding energy of the STT boson stars.

The binding energy thus features a region with three distinct sets of solutions, one GR branch and two STT branches.
The STT branch with the lower binding energy is unstable, but the STT branch with the higher binding energy is at least metastable like the GR branch.
Below the point of equal binding energy, the GR branch is energetically favored, while above this point it is the STT branch that is favored.
The configurations at the point of equal binding energy are completely distinct, though.
Consequently, the phase transition from GR to STT solutions is not smooth but of first order.

We have observed the same pattern when varying the STT parameter $\beta$, except that for large values of $\beta$ the scalarized solutions are never of physical interest, forming only unstable branches.
Similarly, variation of the self-coupling constant does not change the picture.
Inclusion of a small mass of the gravitational scalar, as needed for observational reasons, does not lead to significant changes either.
Considering moreover the presence of rotation also only reinforces the above scenario, and so does inspection of our previous calculations of rotating solutions with higher angular momenta, i.e., higher $n$.
Thus we conjecture that for boson stars with a repulsive potential and mini-boson stars the phase transition is of first order.

\begin{figure}[t!]
\begin{center}
\vspace{-0.5cm}
\mbox{
(a)\hspace*{-0.5cm}\includegraphics[height=.225\textheight, angle =0]{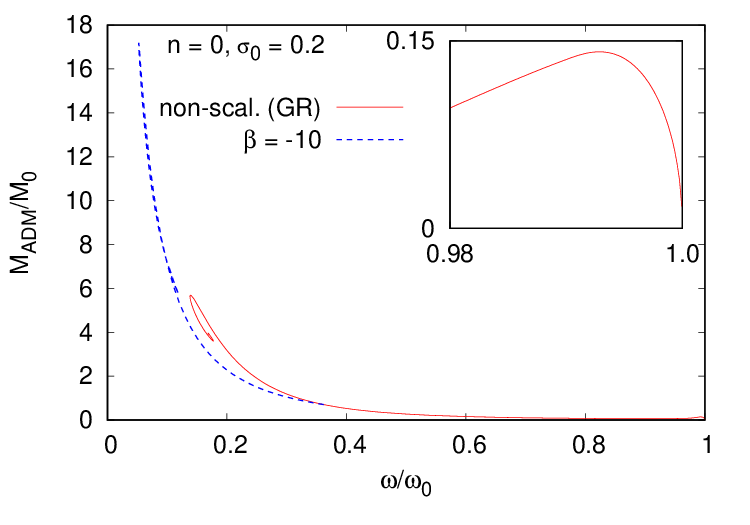}
(b)\hspace*{-0.5cm}\includegraphics[height=.225\textheight, angle =0]{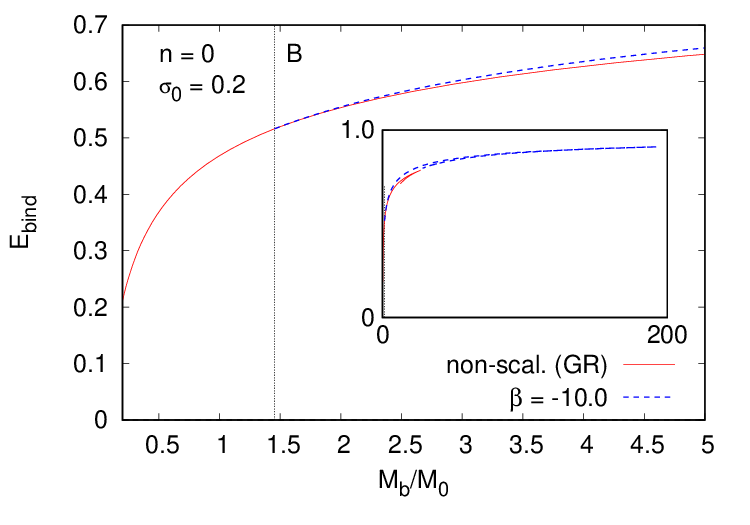}
}
\end{center}
\vspace{-0.5cm}
\caption{Non-rotating solitonic boson stars ($n=0$): 
(a) The scaled ADM mass $M_\text{ADM}/M_0$ versus the scaled boson frequency $\omega/\omega_0$ is shown for non-scalarized (GR) boson stars (solid) with self-interaction coupling constant $\sigma_0=0.2$,
and for scalarized boson stars for a massless ($\beta=-10$, dotted) gravitational scalar field; 
(b) analogous to (a) for the relative binding energy $E_\text{bind}$ versus the scaled boson mass $M_b/M_0$.
The figure shows an enlargement of the region close to the bifurcation, while the inset shows the full region.
\hh{
The thin vertical line indicates the transition point B.}
\label{fig6}
}
\end{figure}

The situation is 
different, though, for other types of self-interactions, like the solitonic self-interaction featuring a potential of sixth order in the boson field \cite{Kleihaus:2005me,Kleihaus:2007vk,Kleihaus:2011sx}.
Spontaneous scalarization for solitonic self-interactions has recently been studied by 
Evstafyeva et al.~\cite{Evstafyeva:2023kfg}, 
employing the  solitonic potential 
\begin{equation}
    U(|\Psi|) = m_b^2 \,|\Psi|^2 
\left( 1 - 2 \frac{|\Psi|^2}{\sigma^2_0} \right)^2 \ ,
\end{equation}
(with $\sigma_0>0$ an input parameter).
Inspection of their results points already to the presence of second-order phase transitions for certain parameter choices.
This is indeed seen in our exploratory calculations, shown in Fig.~\ref{fig6}.
However, 
calculations for other parameter choices 
indicate that first-order transitions occur.
Clearly, this interesting subject should be pursued further to see which order of the phase transition might be the dominant one.

As shown in \cite{Kleihaus:2015iea}, the Kerr black holes with scalar hair (KBHSH) \cite{Herdeiro:2014goa}, that are present in GR, can also have spontaneously scalarized brethren in STTs.
So far, their domain of existence has been mapped out for a small set of parameters.
It appears to be of interest to address the nature of the phase transition also for the KBHSH in GR and STTs.
KBHSH exist for various types of self-interaction potentials \cite{Herdeiro:2014goa,Kleihaus:2015iea,Herdeiro:2015gia,Herdeiro:2015tia}.
Here, an intriguing question would be whether the order of the phase transition carries over from boson stars to their associated  KBHSH.

\hh{Although we have seen that some of the scalarized boson stars obtained in this work are energetically favored, their direct detection remains uncertain. Present constraints from solar-system and pulsar-timing tests already limit the scalar coupling strength and the scalar-field mass, leaving only a narrow parameter window for strong scalarization~\cite{Will:2018bme,Doneva:2022ewd}. 
Future observations, however, could probe this regime. Gravitational-wave and fast radio burst observations may provide complementary probes of boson stars, as their mergers or lensing signatures could reveal characteristic imprints distinct from black holes~\cite{Choi:2019mva}. 
The space-based LISA mission will access lower frequencies and could detect long-lived binaries containing ultralight scalar components~\cite{LISA:2022kgy}. 
On larger scales, the Event Horizon Telescope might test horizonless compact objects whose shadows or photon rings deviate slightly from black-hole predictions~\cite{Olivares:2018abq}. 
Although challenging, these forthcoming multi-messenger observations offer the most promising avenue to constrain—or possibly detect—energetically favored scalarized boson stars in scalar–tensor gravity.}

\section*{Acknowledgment}
%
H. H., M.-Y. L. and D.-C. Z. gratefully acknowledge support by the
National Natural Science Foundation of China (Grant Nos. 12205123,12565010, 12305064, 1236009). 
E. R. gratefully acknowledges the support of the Alexander von Humboldt Foundation.
The work of E. R. is also supported by CIDMA under the FCT Multi-Annual Financing
Program for R\&D Units (UID/04106),
through the Portuguese Foundation for Science and Technology (FCT -- Fundac\~ao para a Ci\^encia e a Tecnologia), as well as the projects:
Horizon Europe staff exchange (SE) programme HORIZON-MSCA2021-SE-01 Grant No. NewFunFiCO-101086251;  
2022.04560.PTDC 
and 2024.05617.CERN. 

\section*{Appendix: Notes on the radius of boson stars} 

\begin{figure}[t!]
\begin{center}
\vspace{-0.5cm}
\mbox{
\includegraphics[height=.225\textheight, angle =0]{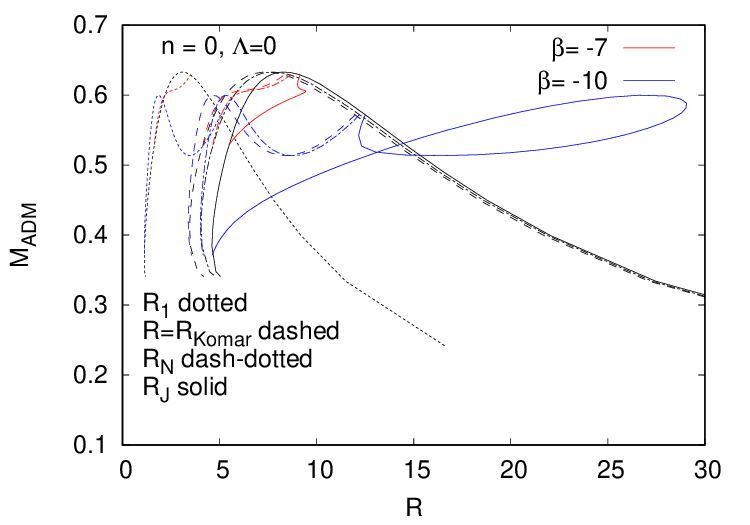}
}
\end{center}
\vspace{-0.5cm}
\caption{Non-rotating mini boson stars ($n=0$, $\Lambda=0$): 
The scaled ADM mass $M_\text{ADM}/M_0$ versus the scaled boson star radius $R$ for 4 definitions of the radius (4 different line styles)
is shown for non-scalarized (GR) boson stars (black) and scalarized boson stars with $\beta=-10$ (blue) and $\beta=-7$ (red).
\label{fig7}
}
\end{figure}

The radius of a boson star is not uniquely defined, since it does not feature a sharp surface.
Consequently, various definitions of the radius have been employed in the literature.
In Fig.~\ref{fig7} we compare a set of definitions for several families of boson stars to demonstrate the widely varying results, particularly for scalarized boson stars.

The definitions employed for calculating the radii are as follows:
\begin{equation}
    R_1= \int_{\Sigma} \sqrt{ -g} \sqrt{\frac{f_1}{f_0}} {\cal A}(\phi) r j^0 d\bar{r} d\theta d\varphi / N \ ,
\end{equation}
where $N$ is the particle number.

Defining the Komar mass function as
\begin{equation}
M_\text{Komar}(r)= \int_0^r \left\{\int \sqrt{ -g} \left( 2 T^0_0 - T^\mu_\mu\right)  d\theta d\varphi\right\} dr \ ,
\end{equation}
where $T^\mu_\nu$ is the total stress-energy tensor.
$R_\text{Komar}$ is defined as the areal radius in the Jordan frame where $M_\text{Komar}(r)$ assumes $99\%$ of the 
ADM mass.

Similarly, defining the particle number function as
\begin{equation}
N(r)=  \int_0^r \left\{\int \sqrt{ -g} j^t d\theta d\varphi\right\} dr \ ,
\end{equation}
$R_\text{N}$ is defined as the areal radius in the Jordan frame where $N(r)$ assumes $99\%$ of the 
particle number.

$R_\text{J}$ in \cite{Evstafyeva:2023kfg} is derived from a mass function given in terms of the metric functions for spherically symmetric boson stars.
However, the mass function can be rewritten as
\begin{equation}
    m(r)= -\int_0^r T^0_0 r^2 dr \ .
\end{equation}
$R_\text{J}$ is defined as the areal radius in the Jordan frame where $m(r)$ assumes $99\%$ of the ADM mass. 
Note that $T^0_0$ contains long range terms like $r^2 {\phi'}^2$ which lead to large values of $R_\text{J}$.
In contrast, $j^0$ and $2 T^0_0 - T^\mu_\mu$ decay exponentially.

We note that for GR boson stars the radii $R_\text{Komar}$, $R_\text{N}$ and $R_\text{J}$ yield very similar values before the maximal mass is reached.
Thus, the resulting values for compactness $C=M/R$ are similar. 
In contrast, the definition $R_1$ yields much smaller values for the radii and thus much larger compactness.
For the scalarized boson stars only $R_\text{Komar}$ and $R_\text{N}$ yield similar values.
Here $R_\text{J}$ shows a very different behavior by producing very large radii for scalarized boson stars in the vicinity of the maximum mass and consequently small values for the compactness.

\end{document}